\def\editmode{0}

\def\bibfilenames{shared_refs}

\def\spsformat{1}

\if\spsformat1

\documentclass{article}
\usepackage{spconf}

\else
\documentclass[11pt,final,onecolumn]{IEEEtran}
\fi

\if\editmode1  
\usepackage[backend=bibtex,style=alphabetic,sorting=debug,maxbibnames=99]{biblatex}
\DeclareFieldFormat{labelalpha}{\thefield{entrykey}}
\DeclareFieldFormat{extraalpha}{}
\bibliography{\bibfilenames}
\newcommand{\cmt}[1]{\noindent\textcolor{lightgreen}{\underline{[#1]}}} 
\newcommand{\hc}[1]{\textcolor{blue}{#1}} 
\newenvironment{myitemize}{\begin{itemize}}{\end{itemize}}
\newcommand{\myitem}{\item}

\else
\usepackage{cite}
\newcommand{\cmt}[1]{} 
\newcommand{\hc}[1]{{#1}} 
\newenvironment{myitemize}{}{}
\newcommand{\myitem}{}

\fi

\newcommand{\printmybibliography}{\if\editmode1 
\printbibliography
\else
\bibliography{\bibfilenames}
\fi
}

\usepackage{fixltx2e}

\usepackage{graphicx}

\usepackage{subcaption}              

\usepackage[utf8]{inputenc}

\usepackage{amsfonts}
\usepackage{amsmath}
\usepackage{mathtools}

\usepackage{amssymb}

\usepackage{bm}
\usepackage{bbm}

\usepackage{color,verbatim}
\usepackage{multirow}
\usepackage{accents}

\usepackage{theoremref}

\usepackage{url}

\newcounter{rulecounter}
\newcommand{\resetrule}{ \setcounter{rulecounter}{0}}
\resetrule

\newsavebox{\selvestebox}
\newenvironment{colbox}[1]
  {\newcommand\colboxcolor{#1}%
   \begin{lrbox}{\selvestebox}%
   \begin{minipage}{\dimexpr\columnwidth-2\fboxsep\relax}}
  {\end{minipage}\end{lrbox}%
   \begin{center}
   \colorbox{\colboxcolor}{\usebox{\selvestebox}}
   \end{center}}

\definecolor{orange}{rgb}{1,0.8,0}
\definecolor{gray}{rgb}{.9,0.9,0.9}
\definecolor{darkgray}{rgb}{.3,0.3,0.3}
\definecolor{darkblue}{rgb}{.1,0.0,0.3}
\definecolor{lightblue}{rgb}{0.7,0.7,1}
\definecolor{lightred}{rgb}{1,0.7,.7}
\definecolor{purple}{RGB}{204,153,255}
\definecolor{lightgray}{rgb}{.95,0.95,0.95}
\definecolor{lightgreen}{rgb}{0.3,0.5,0.3}
\definecolor{darkgreen}{rgb}{0.05,0.3,0.05}

\newcommand{\ra}{$\rightarrow$~}

\newcommand{\acom}[1]{\textcolor{red}{\textbf{[#1]}}}

\newcommand{\brackets}[1]{#1}

\newcommand{\tbm}[1]{{\tilde{\bm #1}}}

\newcommand{\inv}{^{-1}}

\newcommand{\rfield}{\mathbb{R}}

\newcommand{\transpose}{^\top}
 \newcommand{\define}{\triangleq}

\newcommand{\prob}{{\mathbb{P}} }
\newcommand{\expected}{\mathop{\mathbb{E}} }
\newcommand{\cov}{\mathop{\textrm{Cov}}}

\newcommand{\var}[1]{\mathop{\textrm{Var}}\brackets{#1} }
\newcommand{\normal}{\mathcal{N}}

\newtheorem{myproposition}{Proposition}
\newtheorem{myremark}{Remark}
\newtheorem{myproblemstatement}{Problem Statement}
\newtheorem{mylemma}{Lemma}
\newtheorem{mytheorem}{Theorem}
\newtheorem{mydefinition}{Definition}
\newtheorem{mycorollary}{Corollary}

\newcommand{\region}{\hc{\mathcal{X}}}
\newcommand{\regiondim}{{\hc{{d}}}}
\newcommand{\loc}{\hc{\bm x}}
\newcommand{\locmat}{\hc{\bm X}}
\newcommand{\txloc}{\loc_\text{Tx}}

\newcommand{\covmat}{{\hc{\bm{C}}}} 
\newcommand{\mean}{{\hc{{\mu}}}} 
\newcommand{\meanvec}{{\hc{\bm{\mu}}}} 
\newcommand{\updatelinearvec}{{\hc{\bm{a}}}} 
\newcommand{\updatelinearoffset}{{\hc{{b}}}} 
\newcommand{\measlikelihoodvar}{{\hc{{\lambda}}}} 

\newcommand{\dist}{\hc{\delta}}

\newcommand{\pow}{\hc{r}}
\newcommand{\minpow}{\hc{\pow}_{\min}}
\newcommand{\powvec}{\hc{\bm\pow}}
\newcommand{\gridpowvec}{\hc{\powvec}^{\grid}}
\newcommand{\txpow}{\hc{P}_\text{Tx}}

\newcommand{\fsgain}{{\hc{\bar l}}} 
\newcommand{\basepow}{{\hc{l}}} %
\newcommand{\basepowvec}{{\bm \basepow}} %
\newcommand{\shad}{{\hc{ s}}} 
\newcommand{\shadvec}{{\bm \shad}} 
\newcommand{\ushad}{{\hc{\bar \shad}}} 
\newcommand{\ushadmean}{\mean_{\hc{\bar \shad}}} 
\newcommand{\ushadvar}{{\hc{\sigma^2_{\hc{ \shad}}}}} 
\newcommand{\shadcovfun}{{\hc{c}}} 
\newcommand{\fad}{{\hc{w}}} 
\newcommand{\fadvec}{{\bm \fad}} %
\newcommand{\fadvar}{{\hc{\sigma^2_{\fad}}}} 

\newcommand{\tind}{{\hc{t}}} 
\newcommand{\auxtind}{{\hc{\tau}}} 
\newcommand{\tnot}[1]{_{#1}} 
\newcommand{\tfun}{{\hc{T}}} 
\newcommand{\tupdate}{{\hc{t}_\text{upd}}} 

\newcommand{\chind}{{\hc{s}}} 
\newcommand{\chnum}{{\hc{S}}} 
\newcommand{\chnot}[1]{_{#1}} 

\newcommand{\measpow}{\tilde{\pow}}
\newcommand{\measpowvec}{{\tbm{\pow}}}
\newcommand{\measnoise}{{\hc{z}}}
\newcommand{\measnoisevec}{\bm{\measnoise}}
\newcommand{\measnoisevar}{{\hc{\sigma^2_{\measnoise}}}} 

\newcommand{\serv}{{\hc{\beta}}} 
\newcommand{\servprob}{\hc{p}^{\serv}} 
\newcommand{\servvec}{{\hc{\bm \serv}}} 
\newcommand{\gridservvec}{\hc{\servvec}^{\grid}}

\newcommand{\grid}{\hc{\mathcal{G}}}
\newcommand{\gridnum}{{\hc{{G}}}} 
\newcommand{\gridind}{{\hc{{g}}}} 
\newcommand{\gridnot}[1]{_{#1}} 
\newcommand{\gridloc}{{\loc}^{\grid}}

\newcommand{\gridbasepowvec}{\basepowvec^{\grid}} 
\newcommand{\gridshadvec}{\shadvec^{\grid}}
\newcommand{\gridfadvec}{\fadvec^{\grid}}

\newcommand{\uncert}{\hc{{u}}} 
\newcommand{\uncertvec}{\hc{\bm{\uncert}}} 

\newcommand{\policyfun}{\hc{{\pi}}} 
\newcommand{\discount}{\hc{{\gamma}}} 

\renewcommand{\paragraph}[1]{\textbf{#1.}}
\newcommand{\nextv}[1]{} 
\newcommand{\thisv}[1]{#1} 

\begin{document}

\title{Aerial Spectrum Surveying: \\Radio Map Estimation with Autonomous UAVs}

\if\spsformat1
\name{
Daniel Romero$^1$, Raju Shrestha$^1$, Yves Teganya$^1$, and Sundeep Prabhakar Chepuri$^2$ \thanks{
Research funded by the Research Council of Norway (IKTPLUSS
grant 280835) and the Indian Department of Science and Technology. \{daniel.romero,raju.shrestha, yves.teganya\}@uia.no, spchepuri@iisc.ac.in.}}
\address{
$^1$Department of Information and Communication Technology, University
of Agder, Norway.\\
$^2$Department of Electrical Communication Engineering, Indian
Institute of Science, India.
}
\else
\author{Daniel Romero$^1$, Raju Shrestha$^1$, Yves Teganya$^1$, and Sundeep Prabhakar Chepuri$^2$ \thanks{
Research has been funded by the Research Council of Norway (IKTPLUSS
grant 280835) and the Indian Department of Science and Technology.}}
\fi

\maketitle

\begin{abstract}
Radio maps are emerging as a popular means to endow next-generation
wireless communications with situational awareness. In particular,
radio maps are expected to play a central role in unmanned aerial
vehicle (UAV) communications since they can be used to determine
interference or channel gain at a spatial location where a UAV has
not been before.  Existing methods for radio map estimation utilize
measurements collected by sensors whose locations cannot be
controlled. In contrast, this paper proposes a scheme in which a UAV
collects measurements along a trajectory. This trajectory is designed
to obtain accurate estimates of the target radio map in a short time
operation. The route planning algorithm relies on a map uncertainty
metric to collect measurements at those locations where they are more
informative. An online Bayesian
learning algorithm is developed to update the map estimate and
uncertainty metric every time a new measurement is collected,
which enables real-time operation.
\end{abstract}


\section{Introduction}

\cmt{motivation}%
\begin{myitemize}%
\myitem\cmt{motivation radio maps \ra situational awareness}%
\begin{myitemize}%
\myitem\cmt{In general}Radio maps find a myriad of applications
in wireless communications, such as network planning, interference
coordination, power control, spectrum management, resource allocation,
handoff procedure design, dynamic
spectrum access, and cognitive
radio; see e.g.~\cite{grimoud2010rem,yilmaz2013radio\nextv{,dallanese2011powercontrol}}.
\nextv{Esp. usage of shorter wavelengths \ra line of
sight...}%
\myitem\cmt{Aerial communications}Recently, radio maps have received 
great attention for autonomous UAV communications and operations; see
e.g.~\cite{chen2017map\nextv{,zhang2019path,chen2018mapassisted},zhang2019constrained}.
\begin{myitemize}%
\myitem\cmt{line of sight \ra interference}%
\myitem\cmt{command and control..., telemetry}%
\myitem\cmt{mission critical}%
\myitem\cmt{UAV-enabled comms}%
\end{myitemize}%
\end{myitemize}%
\myitem\cmt{motivation surveying}These observations call for the
development of a technology for ``surveying'' a spatial region of
interest to construct a radio map.
\myitem\cmt{goal: surveying}The goal of this paper is to address this task by collecting
measurements  with an autonomous UAV.
\end{myitemize}

\cmt{literature}
\begin{myitemize}
\myitem\cmt{given locations}Over the last few years, a significant body of
literature has addressed the estimation of radio maps from
measurements acquired by spatially distributed sensors, typically by
some form of interpolation
algorithm. This includes
\begin{myitemize}%
\myitem\cmt{general cartograpy}%
\begin{myitemize}%
\myitem\cmt{kriging}kriging\cite{alayafeki2008cartography}\nextv{\cite{alayafeki2008cartography,boccolini2012wireless, agarwal2018spectrum}},
\myitem\cmt{compress. sensing, dict. lear., matrix
completion}compressed
sensing~\cite{jayawickrama2013compressive,bazerque2010sparsity},
dictionary learning~\cite{\nextv{kim2011link,}kim2013dictionary},
matrix~\cite{\nextv{ding2016cellular,}lee2016lowrank} and
\myitem\cmt{tensor compl, decomp}tensor
completion~\cite{\nextv{zhang2019spectrum,}tang2016spectrum},
\myitem\cmt{bayes. models, radial basis functions, and kernel
methods}Bayesian models~\cite{huang2015cooperative\nextv{,lee2019bayesian}}, \nextv{radial basis functions~\cite{hamid2017non, zha2018spectrum},} kernel
methods~\cite{teganya2019locationfree\nextv{,bazerque2013basispursuit},romero2017spectrummaps,romero2018blind},
thin-plate spline
regression~\cite{bazerque2011splines}, 
and 
\myitem\cmt{deep learning}deep
learning~\cite{han2020power,teganya2020autoencoders}.
\end{myitemize}%
\myitem\cmt{with UAVs}In the  context of UAV communications,
\begin{myitemize}
\myitem\cmt{segmented}radio map estimators have been proposed in~\cite{chen2017segmented}. 
\end{myitemize}%
\myitem\cmt{Limitations}All these schemes assume that the measurement positions
are given and, hence, cannot decide where to measure
next. 
\end{myitemize}%
\myitem\cmt{simultaneous}Another related  scheme is the one in \cite{zheng2020simultaneous},
which does decide the trajectory of a UAV. However, the criterion is to
 minimize an outage metric and, thus, not tailored to construct a
 radio map.

\nextv{
\myitem\cmt{further related literature?}
\begin{myitemize}
\myitem\cmt{design of experiments}\acom{Sundeep, can you provide a
general reference?}
\myitem\cmt{robot exploration/mobile sensing}The problem in this paper
falls also in the area of robot exploration and mobile sensing; see
e.g.~\cite{leny2009trajectory,krause2007submodular}. However, the
focus of this literature typically pertains to \emph{simultaneous
localization and mapping} and computer vision~\acom{cite}. To the best
of our knowledge, no existing scheme is capable of adaptively
determining measurement trajectories with the uncertainty metrics
adopted here and at a low computational complexity, as required by UAV
platforms. 
\end{myitemize}
}
\end{myitemize}

\cmt{contributions}This paper fills this gap by proposing \emph{aerial
spectrum surveying}, whereby a UAV autonomously collects measurements
 across the area of interest and adaptively decides where to measure
 next so that the time required to attain a prescribed estimation accuracy
 is approximately minimized.\footnote{Although the focus is on UAVs,
 most of the ideas here can be extended to other mobile robots such as
 terrestrial vehicles.} To this end, the following challenges
 are addressed:
\begin{myitemize}%
\myitem\cmt{online Bayesian learning}(i) Since there are infinitely many candidate
 measurement locations in 3D space, the UAV needs to judiciously
 select an informative finite subset of them. To this end, a Bayesian
 learning scheme is adopted to estimate the radio map along with its
 uncertainty across space. Since adaptively planning the trajectory
 requires updating this uncertainty metric as more measurements are
 collected, an online learning algorithm with constant complexity per
 measurement is developed.
\myitem\cmt{Route planning}(ii) Given the aforementioned 
 metric, the UAV needs to plan a trajectory that prioritizes those
 points with a high uncertainty. To cope with the combinatorial
 complexity involved in this kind of formulations, two approximations
 are explored. The first relies on a receding horizon formulation cast
 as a \emph{discounted-reward} travelling salesman problem, for which
 polynomial complexity approximations
 exist~\cite{blum2007orienteering}.  Since this complexity may still
 be unaffordable for real-time operation on board an UAV, a simpler
 waypoint-search scheme based on a shortest-path subroutine and a
 suitably designed spatial cost matrix is devised. This approach
 provides measurement locations at a low complexity while accounting
 for uncertainty and experience. The price to be paid is an increased
 suboptimality.

\end{myitemize}

\cmt{paper structure}Sec.~\ref{sec:estimation}  addresses the
 contributions in (i) whereas Sec.~\ref{sec:trajectory} addresses those
 in (ii). The proposed scheme is validated through simulations in
 Sec.~\ref{sec:experiments}.

\cmt{Notation}\emph{Notation:} Boldface lowercase (uppercase) denote
 column vectors (matrices). For a random vector $\bm x$, notation
 $\normal(\bm x|\bm \mu, \bm C)$ or, its short-handed version
 $\normal(\bm \mu, \bm C)$, denotes a Gaussian distribution with mean
 $\bm \mu$ and covariance matrix $\bm C$.

\section{Online  Radio Map Learning}
\label{sec:estimation}

\cmt{overview}After presenting the model, this section formulates the
 problems of estimating \emph{power} and \emph{service maps} as well
 as their associated uncertainty.

\subsection{Radio Map Model}
\label{sec:model}

\cmt{space}
\begin{myitemize}
\myitem\cmt{region}Let $\region \subset \rfield ^\regiondim$ represent
the geographical region of interest, where $\regiondim$ is either 2 or 3,
\myitem\cmt{tx location}and consider a transmitter at location
$\txloc\in \region$. This transmitter may correspond to a cellular
base station. The location
$\txloc$
\myitem\cmt{tx pow}as well as the transmit power $\txpow$ can be
assumed known as base stations in contemporary cellular networks
share this information with the users.
\cmt{single-channel first}A single transmitter is assumed to keep the
notation simple, but multiple transmitters can be readily
accommodated. 
\end{myitemize}%
\cmt{rx power}%
\begin{myitemize}%
\myitem\cmt{rx signal power at $\loc\in\region$}As usual, the power received  at
$\loc\in\region$ is given in logarithmic units by
\begin{align}
\label{eq:explicitmodel}
\pow(\loc) = \txpow + \fsgain(\loc) - \ushad(\loc) + \fad(\loc)
\end{align}
where  each term is explained next.
\begin{myitemize}%
\myitem\cmt{path loss}$\fsgain(\loc)$ captures free-space
path loss \nextv{between $\txloc$ and $\loc$} and antenna
gain. 
\myitem\cmt{shadowing}$\ushad(\loc)$ is the shadowing loss, which
 captures attenuation due to obstructions.
\begin{myitemize}%
\myitem\cmt{Gaussian}With the usual log-normal 
assumption, let $\ushad(\loc)\sim\normal(\ushadmean,\ushadvar)$.
\myitem\cmt{correlation}\nextv{Regarding its spatial correlation, a customary
assumption (see e.g. \acom{cite also papers on kriging}) is that
$\cov(\ushad(\loc),\ushad(\loc'))$ depends only on the distance, i.e.,
$\cov(\ushad(\loc),\ushad(\loc'))=\shadcovfun(||\loc-\loc'||^2)$. In~\cite{gudmundson1991correlation},
an explicit form for function $\shadcovfun(\cdot)$ is obtained based
on a measurement campaign. To facilitate its interpretability, this
expression will be reparameterized here as
$\shadcovfun(\dist)=\ushadvar 2^{-\dist/\dist_0}$, where $\dist_0$ is
the distance at which the correlation decays to 1/2.}
\thisv{Following the empirical model
in~\cite{gudmundson1991correlation},
$\cov(\ushad(\loc),\ushad(\loc'))=\shadcovfun(||\loc-\loc'||)$,
where function $\shadcovfun$ is reparameterized here as
$\shadcovfun(\dist)=\ushadvar 2^{-\dist/\dist_0}$ with $\dist_0$ 
the distance at which the correlation decays to 1/2.}
\end{myitemize}%
\myitem\cmt{fading}Finally,  $\fad(\loc)$ accounts for small-scale
fading\nextv{ between $\txloc$ and $\loc$}, caused by the
constructive/destructive interference between the signal paths
arriving at $\loc$, \nextv{. For wavelengths used in contemporary
communications, this changes in the order of centimeters. Thus, there
is little hope to estimate this term using an UAV given the
positioning error with existing global navigation satellite systems
(GNSSs). However, the measurement process itself tends to average out
this term across space when the sensor moves, as is the case here.}%
\nextv{
\begin{myitemize}
\myitem\cmt{mean}zero-mean perturbation with
\myitem\cmt{var}variance $\fadvar$
\myitem\cmt{indep}and independent of $\fad(\loc')$ and
$\ushad(\loc'')$  for all $ \loc', \loc''\in \region$ with $\loc'\neq\loc$. 
\myitem\cmt{unmodeled}Further unmodeled effects may also be  absorbed
in this term, which motivates the additional assumption that $\fad(\loc)$ is 
Gaussian distributed. 
\end{myitemize}
}%
\thisv{
as well as additional unmodeled effects. As
in~\cite{huang2015cooperative}, $\fad(\loc)$ will be modeled as
$\normal( 0, \fadvar)$. Additionally, it is assumed independent of $\fad(\loc')$
and $\ushad(\loc'')$ for all $ \loc', \loc''\in \region$ with
$\loc'\neq\loc$.  }
\end{myitemize}%
\myitem\cmt{Simplified notation}For clarity, rewrite \eqref{eq:explicitmodel} as
\begin{align}
\label{eq:simplifiedmodel}
\pow(\loc) = \basepow(\loc) - \shad(\loc) + \fad(\loc),
\end{align}
where $\basepow(\loc)\define\txpow+\fsgain(\loc)-\ushadmean$ and
$\shad(\loc)\define \ushad(\loc)-\ushadmean$. The deterministic
component $\basepow(\loc)$ can be assumed known as $\ushadmean$ can
be readily estimated from a set of measurements.
\end{myitemize}%

\cmt{measurements}To estimate the radio map, a UAV equipped  with a
communication module capable of measuring power and a GPS sensor
collects measurements
\begin{myitemize}%
\myitem\cmt{def}$(\loc\tnot{\auxtind},\measpow\tnot{\auxtind})$,
$\auxtind=0,1,\ldots$, where
$\measpow\tnot{\auxtind}\define \pow(\loc\tnot{\auxtind})
+ \measnoise\tnot{\auxtind}$ is the \emph{received signal strength} at
 $\loc\tnot{\auxtind}\in \region$ and
$\measnoise\tnot{\auxtind}\sim\normal(0,\measnoisevar)$ models the
measurement error, assumed independent across $\auxtind$ and
independent of $\fad(\loc)$ and $\shad(\loc')$ for all
$\loc, \loc'\in \region$.
\myitem\cmt{set}The measurements and their locations up to and including
time $\tind$ will be arranged as $\measpowvec\tnot{\tind}\define
[\measpow\tnot{0},\ldots,\measpow\tnot{\tind}]\transpose \in \rfield
^{\tind+1}$ and
$\locmat\tnot{\tind}\define[\loc\tnot{0},\ldots,\loc\tnot{\tind}]\in \rfield^{\regiondim
\times (\tind+1)}$.
\end{myitemize}%

\subsection{Estimation Problem Formulation}
\cmt{overview}This section formulates the problem of estimating
two classes of fradio maps given a collection of measurements.

\paragraph{Power Map Estimation}
\label{sec:powerestimationproblem}
\cmt{grid notation}%
\begin{myitemize}%
\myitem\cmt{motivation}Given the above model, the \emph{power map}
$\pow(\loc)$ can be estimated with a conventional Gaussian-process
estimator~\cite[Sec. 6.4]{bishop2006}.
\nextv{The model introduced in Sec.~\ref{sec:model}
fits the framework of Gaussian processes \acom{cite Bishop} and one
could therefore estimate function $\pow(\loc)$ along the lines
of \acom{cite cartography Bayesian}.}Unfortunately,
such \emph{non-parametric} approaches incur unbounded complexity as
their estimates involve the summation of one term per data point.
\nextv{Thus,
these approaches would eventually become computational prohibitive. }%
\myitem\cmt{def}To circumvent this effect, a key idea here
is to aggregate the information provided by all the measurements up to
and including time $\tind$ by the posterior of $\pow(\loc)$ at a finite set of
arbitrary grid
points
$\grid\define\{\gridloc\gridnot{0},\ldots,\gridloc\gridnot{\gridnum-1}\}\subset \region$\nextv{of
$\gridnum$ distinct points. These points can be arbitrarily located
across $\region$, not necessarily in a regular arrangement. For
example, more grid points may be placed at areas with heavier traffic.}.
\myitem\cmt{power values on the grid}At these points, let (cf. \eqref{eq:simplifiedmodel})
\begin{align}
\label{eq:gridpowvec}
\gridpowvec\define[\pow(\gridloc\gridnot{0}),\ldots,\pow(\gridloc\gridnot{\gridnum-1})]\transpose
= \gridbasepowvec -\gridshadvec + \gridfadvec
\end{align}
where
\begin{myitemize}%
\myitem\cmt{}$\gridbasepowvec\define[\basepow(\gridloc\gridnot{0}),\ldots,\basepow(\gridloc\gridnot{\gridnum-1})]\transpose$, \myitem\cmt{}$\gridshadvec\define[\shad(\gridloc\gridnot{0}),\ldots,$ $\shad(\gridloc\gridnot{\gridnum-1})]\transpose$,
and
 \myitem\cmt{}$\gridfadvec\define[\fad(\gridloc\gridnot{0}),\ldots,\fad(\gridloc\gridnot{\gridnum-1})]\transpose$.
\end{myitemize}%
\end{myitemize}%
\cmt{batch problem}The \emph{batch} version of the problem
is 
\begin{myitemize}%
\myitem\cmt{formulation}
\begin{myitemize}%
\myitem\cmt{find}to obtain  $p(\gridpowvec|\measpowvec\tnot{\tind},\locmat\tnot{\tind})$
\myitem\cmt{given}given $\measpowvec\tnot{\tind}$ and $\locmat\tnot{\tind}$.
\end{myitemize}%
\myitem\cmt{Interpretation}%
\nextv{
\begin{myitemize}%
\myitem\cmt{Mean}Applying a  well-known result in estimation theory \acom{cite
kay}, the mean of such a posterior distribution provides
the \emph{minimum mean square error} (MMSE) estimate of $\pow(\loc)$
at the grid points.
\myitem\cmt{covariance}On the other hand, the covariance of this posterior captures the uncertainty
in one's knowledge of the true $\gridpowvec$ after observing the
measurements. Although this can indeed be utilized to determine the UAV
trajectory, this formulation still suffers from a limitation: if
all measurements are processed at once every time a new posterior
needs to be computed, the complexity will grow superlinearly with
$\tind$ and therefore will become unaffordable.
\end{myitemize}%
}\thisv{One can then retrieve an estimate of $\gridpowvec$ as the mean of this
posterior and an uncertainty metric from the covariance.}
\end{myitemize}%
\cmt{online problem}However, given the unbounded complexity that such
a task may entail, it is more convenient to address the \emph{online}
problem of iteratively\nextv{estimation task, which means that each new measurement is utilized to
refine the previous posterior.}
\begin{myitemize}%
\myitem\cmt{formulation}\nextv{Specifically, consider the problem of }%
\begin{myitemize}%
\myitem\cmt{find}finding $p(\gridpowvec|\measpowvec\tnot{\tind},\locmat\tnot{\tind})$
\myitem\cmt{given}given the previous posterior
$p(\gridpowvec|\measpowvec\tnot{\tind-1},\locmat\tnot{\tind-1})$ and
the most recent measurement
$(\loc\tnot{\tind},\measpow\tnot{\tind})$ with bounded complexity per $\tind$.
\end{myitemize}%
\myitem\cmt{interpretation}\nextv{Imposing that the complexity of this
operation must not grow with $\tind$ yields a viable approach to
update the state-of-knowledge of the estimator with an unlimited
number of observations given the available computational resources.}
\end{myitemize}%

\nextv{
\cmt{Multiple transmitters}So far,  a single
transmitter has been considered. In practice, the same model carries
over to scenarios where each transmitter operates on a different
frequency channel or, simply, whenever the sensor can discern the
power received from each transmitter, e.g. because different
transmitters use different spreading sequences. In that case, one
would have a map $\pow\chnot{\chind}(\loc)$ per transmitter (or
source) $\chind=0,1,\ldots,\chnum-1$ and  corresponding
equations \eqref{eq:simplifiedmodel} and \eqref{eq:gridpowvec}. 
On the contrary, if the sensor cannot distinguish transmitters, the
same model can be extended along the lines of \acom{cite kriging} and
the ensuing derivations carry over. In case that there is overlap
across frequency, the present approach can be extended to accommodate
a \emph{basis expansion model}; see \acom{cite psd cartography
papers}. However, most of this paper will focus on the presented
simplified model for the sake of readability. 
}
\cmt{Time variation?}

\paragraph{Service Map Estimation}
\label{sec:serviceestimationproblem}%
\begin{myitemize}%
\myitem\cmt{Motivation service probability}In UAV applications, rather
than knowing the exact value of $\pow(\loc)$, it is often more
relevant to know the set of locations $\loc$ that the base station can
serve with a prescribed binary rate. 
\begin{myitemize}%
\myitem\cmt{UAV command and control channel}This is necessary e.g. to
establish a command-and-control channel or to communicate
application-dependent data.
\myitem\cmt{extension to SNR}%
\end{myitemize}%
\myitem\cmt{service def}Since the scheme can be readily extended to
accommodate interference, assume for simplicity that the throughput is
limited by noise and, therefore, one can regard location $\loc$ as
served if $\pow(\loc)\geq \minpow$ for a given $\minpow$. Let
$\serv(\loc)=1$ in that case and  $\serv(\loc)=0$ otherwise.
\myitem\cmt{focus on grid}\nextv{ $\gridservvec\define[\serv(\gridloc\gridnot{0}),\ldots,\serv(\gridloc\gridnot{\gridnum-1})]\transpose$}%
\myitem\cmt{problem}The  problem in this case is to find
$p(\serv(\gridloc\gridnot{\gridind})|\measpowvec\tnot{\tind},\locmat\tnot{\tind})$ for
each $\gridind$.  The online and batch versions can be phrased as
before. 
\end{myitemize}%

\subsection{Batch and Online Bayesian Estimators}

\cmt{overview}Although the focus is on online learning, the  solution
to the batch problem is briefly described first to facilitate
understanding.
\cmt{notation}For notational convenience, let
\begin{align}
\label{eq:measpowvec}
\measpowvec\tnot{\tind} = \basepowvec\tnot{\tind} - \shadvec\tnot{\tind} + \fadvec\tnot{\tind} + \measnoisevec\tnot{\tind},
\end{align}
where
\begin{myitemize}%
\myitem\cmt{}$\basepowvec\tnot{\tind}\define[\basepow(\loc\tnot{0}),\ldots,\basepow(\loc\tnot{\tind})]\transpose$,
\myitem\cmt{}$\shadvec\tnot{\tind}\define[\shad(\loc\tnot{0}),\ldots,\shad(\loc\tnot{\tind})]\transpose$,
\myitem\cmt{}$\fadvec\tnot{\tind}\define[\fad(\loc\tnot{0}),\ldots,\fad(\loc\tnot{\tind})]\transpose$, and
\myitem\cmt{}$\measnoisevec\tnot{\tind}\define[\measnoise\tnot{0},\ldots,\measnoise\tnot{\tind}]\transpose$.
\end{myitemize}%

\paragraph{Batch Power Map Estimator}
\label{sec:batchestimator}
\cmt{decomposition}
\cmt{$\gridpowvec \perp \measpowvec\tnot{\tind}
| \gridshadvec$}From the model embodied by  \eqref{eq:gridpowvec}
  and \eqref{eq:measpowvec}, it can be readily shown that 
$\gridpowvec$ is conditionally independent of
  $\measpowvec\tnot{\tind}$ given $\gridshadvec$. This, in turn,
  implies that 
\begin{align}
\label{eq:marginalbatch}
p(\gridpowvec|\measpowvec\tnot{\tind}) = \int p(\gridpowvec|\gridshadvec) p(\gridshadvec|\measpowvec\tnot{\tind})d\gridshadvec,
\end{align}
where $\locmat\tnot{\tind}$ has been omitted to lighten the notation.
\begin{myitemize}%
\myitem\cmt{First term}From \eqref{eq:gridpowvec} and the fact that 
$\gridbasepowvec$ is deterministic, it clearly follows
that the first factor in the integrand is 
 $p(\gridpowvec|\gridshadvec)=\normal(\gridpowvec| \gridbasepowvec
 -\gridshadvec, \fadvar \bm I_\gridnum)$.
\myitem\cmt{Second term}To obtain the  second factor
$p(\gridshadvec|\measpowvec\tnot{\tind})$, observe that $\gridshadvec$
and $\measpowvec\tnot{\tind}$ are jointly Gaussian.
\begin{myitemize}%
\myitem\cmt{joint}
In particular, one
can obtain the parameters of their joint distribution
$p(\gridshadvec,\measpowvec\tnot{\tind})$ as follows. 
\begin{myitemize}%
\myitem\cmt{mean}First, the mean vectors are clearly
$\expected[\gridshadvec]=\bm 0$ and $\expected[\measpowvec\tnot{\tind}]
= \basepowvec\tnot{\tind}$. 
\myitem\cmt{covariance}For the covariance, let $\cov[\gridshadvec]\define\covmat_{
 \gridshadvec}$ and write $\cov[\gridshadvec,\measpowvec\tnot{\tind}]=\expected[
\gridshadvec(\measpowvec\tnot{\tind} -\basepowvec\tnot{\tind}) 
\transpose]
=\expected[
\gridshadvec(- \shadvec\tnot{\tind} + \fadvec\tnot{\tind} + \measnoisevec\tnot{\tind}) 
\transpose]
=-\expected[
\gridshadvec \shadvec\tnot{\tind} \transpose]
 \define-\covmat_{\gridshadvec,
 \shadvec\tnot{\tind}}
$ as well as $\cov[\measpowvec\tnot{\tind}]=\expected[
(\measpowvec\tnot{\tind} -\basepowvec\tnot{\tind})
(\measpowvec\tnot{\tind} -\basepowvec\tnot{\tind})\transpose]
=\expected[
(- \shadvec\tnot{\tind} + \fadvec\tnot{\tind}
+ \measnoisevec\tnot{\tind})
(- \shadvec\tnot{\tind} + \fadvec\tnot{\tind} + \measnoisevec\tnot{\tind})\transpose]
= \cov[\shadvec\tnot{\tind}] + \fadvar\bm I_{\tind+1}
+ \measnoisevar\bm I_{\tind+1}
\define \covmat_{\shadvec\tnot{\tind}} + \fadvar\bm I_{\tind+1}
+ \measnoisevar\bm I_{\tind+1} $. Here, the matrices
$\covmat_{ \gridshadvec}$,
$\covmat_{\gridshadvec, \shadvec\tnot{\tind}}$ and
$\covmat_{\shadvec\tnot{\tind}}$ can be obtained from the covariance
function $\shadcovfun$ introduced in Sec.~\ref{sec:model}.
\end{myitemize}%
\myitem\cmt{shadowing posterior}Applying \cite[Th. 10.2]{kay1} to this
joint distribution, it follows
that $p(\gridshadvec|\measpowvec\tnot{\tind})
= \normal(\gridshadvec|\meanvec_{\gridshadvec|\measpowvec\tnot{\tind}}, \covmat_{\gridshadvec|\measpowvec\tnot{\tind}})$, where
\begin{align*}
\meanvec_{\gridshadvec|\measpowvec\tnot{\tind}}&=
\cov[\gridshadvec,\measpowvec\tnot{\tind}]{\cov}\inv[\measpowvec\tnot{\tind}](\measpowvec\tnot{\tind}-
\expected[\measpowvec\tnot{\tind}])
\\&=
-\covmat_{\gridshadvec,
 \shadvec\tnot{\tind}}
( \covmat_{\shadvec\tnot{\tind}} + \fadvar\bm I_{\tind+1}
+ \measnoisevar\bm I_{\tind+1})\inv
(\measpowvec\tnot{\tind}- \basepowvec\tnot{\tind})
\\
\covmat_{\gridshadvec|\measpowvec\tnot{\tind}}&=
\cov[\gridshadvec]
-\cov[\gridshadvec,\measpowvec\tnot{\tind}]{\cov}\inv[\measpowvec\tnot{\tind}]
\cov[\measpowvec\tnot{\tind},\gridshadvec]
\\&=\covmat_{\gridshadvec}
-\covmat_{\gridshadvec,
 \shadvec\tnot{\tind}}
( \covmat_{\shadvec\tnot{\tind}} + \fadvar\bm I_{\tind+1}
+ \measnoisevar\bm I_{\tind+1})\inv
\covmat_{
 \shadvec\tnot{\tind},\gridshadvec},
\end{align*}
where $\covmat_{ \shadvec\tnot{\tind},\gridshadvec}\define\covmat_{\gridshadvec, \shadvec\tnot{\tind}}\transpose$.
\end{myitemize}%
\end{myitemize}%
\cmt{recomposition}Finally, applying  \cite[eq. (2.115)]{bishop2006}
to obtain the conditional marginal \eqref{eq:marginalbatch} yields
 $p(\gridpowvec|\measpowvec\tnot{\tind}) = \normal(\gridpowvec|
\meanvec_{\gridpowvec|\measpowvec\tnot{\tind}}
,\covmat_{\gridpowvec|\measpowvec\tnot{\tind}}
)$
with $\meanvec_{\gridpowvec|\measpowvec\tnot{\tind}}\define
 \basepowvec\tnot{\tind}-\meanvec_{\gridshadvec|\measpowvec\tnot{\tind}}$
and $\covmat_{\gridpowvec|\measpowvec\tnot{\tind}}\define\fadvar \bm I_\gridnum
+\covmat_{\gridshadvec|\measpowvec\tnot{\tind}}
$, thereby solving the batch problem.

\paragraph{Online Power Map Estimator}
\label{sec:onlineestimator}
\cmt{decomposition}%
\begin{myitemize}%
\myitem\cmt{Goal}To address the online power map estimation problem (see
Sec.~\ref{sec:powerestimationproblem}), it is convenient to decompose
$p(\gridpowvec|\measpowvec\tnot{\tind})$ into \nextv{the previous posterior}$
p(\gridpowvec|\measpowvec\tnot{\tind-1})$ and a term that depends on
the last measurement only. 
\myitem\cmt{conditional indep.}\nextv{As described in
Sec.~\ref{sec:powerestimationproblem},}However, it can be easily seen
that such a factorization is not possible due to the posterior
correlation among measurements. To sidestep this difficulty, the
central idea in the proposed online learning scheme (see also
Sec.~\ref{sec:powerestimationproblem}) is to use $\grid$ to summarize
the information of all past measurements. Mathematically, this can be
phrased as the assumption that $\measpow\tnot{\tind}$ and
$\measpowvec\tnot{\tind-1}$ are conditionally independent given
$\gridpowvec$. That is, when $\gridpowvec$ is known, the past
measurements $\measpowvec\tnot{\tind-1}$ do not provide extra
information about $\measpow\tnot{\tind}$. The error that this
approximation introduces can be reduced by adopting a denser grid and
pays off since it enables online estimation.

\myitem\cmt{bayes}From this assumption and Bayes' rule,  it follows
  that
  \begin{subequations}
\begin{align*}
p(\gridpowvec&|\measpowvec\tnot{\tind})=p(\gridpowvec|\measpow\tnot{\tind},\measpowvec\tnot{\tind-1})
\propto
p(\measpow\tnot{\tind},\measpowvec\tnot{\tind-1}|\gridpowvec)p(\gridpowvec)
\\&=
p(\measpow\tnot{\tind}|\gridpowvec)p(\measpowvec\tnot{\tind-1}|\gridpowvec)p(\gridpowvec)
=
p(\measpowvec\tnot{\tind-1},\gridpowvec)p(\measpow\tnot{\tind}|\gridpowvec)
\\&=
p(\gridpowvec|\measpowvec\tnot{\tind-1})p(\measpowvec\tnot{\tind-1})p(\measpow\tnot{\tind}|\gridpowvec)
\propto
p(\gridpowvec|\measpowvec\tnot{\tind-1})p(\measpow\tnot{\tind}|\gridpowvec),
\end{align*}
\end{subequations}
where $\propto$ denotes equality up to a positive factor\nextv{ scaling constant. In
this context, a constant is understood as a positive term} that does not depend
on $\gridpowvec$.
\begin{myitemize}%
\myitem\cmt{first term}As shown earlier in this section,
$p(\gridpowvec|\measpowvec\tnot{\tind-1})=\normal(\gridpowvec|\meanvec_{\gridpowvec|\measpowvec\tnot{\tind-1}},\covmat_{\gridpowvec|\measpowvec\tnot{\tind-1}})$. Since
 $p(\gridpowvec|\measpowvec\tnot{\tind-1})$ is given  in the online
formulation (cf. Sec.~\ref{sec:powerestimationproblem}), the
online learning algorithm can use 
$\meanvec_{\gridpowvec|\measpowvec\tnot{\tind-1}}$ and
$\covmat_{\gridpowvec|\measpowvec\tnot{\tind-1}}$
to obtain  $p(\gridpowvec|\measpowvec\tnot{\tind})$.

\myitem\cmt{second term}To find
$p(\measpow\tnot{\tind}|\gridpowvec)$, note that $\measpow\tnot{\tind}$ 
and $\gridpowvec$ are jointly Gaussian. It follows
from \cite[Th. 10.2]{kay1} that $p(\measpow\tnot{\tind}|\gridpowvec)$
is Gaussian distributed with parameters
\begin{align*}
\expected&[\measpow\tnot{\tind}|\gridpowvec]
=
\expected[\measpow\tnot{\tind}] +
\cov[\measpow\tnot{\tind},\gridpowvec]
{\cov}\inv[\gridpowvec]
(\gridpowvec-\expected[\gridpowvec])\\
&=
\basepow(\loc\tnot{\tind}) +
\expected[(-\shad(\loc\tnot{\tind})+\fad(\loc\tnot{\tind})+\measnoise\tnot{\tind})
        (-\gridshadvec + \gridfadvec)\transpose]
\\&\quad\times{\expected}\inv[(-\gridshadvec + \gridfadvec)(-\gridshadvec
+ \gridfadvec)\transpose]
(\gridpowvec-\gridbasepowvec)\\
&=
\basepow(\loc\tnot{\tind}) +
(\covmat_{\shad(\loc\tnot{\tind}),\gridshadvec}
+
\covmat_{\fad(\loc\tnot{\tind}),\gridfadvec})
\\&\quad\times(\covmat_{\gridshadvec} + \fadvar \bm I_\gridnum)\inv
(\gridpowvec-\gridbasepowvec)\define \updatelinearvec\tnot{\tind}\transpose \gridpowvec + \updatelinearoffset\tnot{\tind}
\\
&\var[\measpow\tnot{\tind}|\gridpowvec]
=
\var[\measpow\tnot{\tind}]
-
\cov[\measpow\tnot{\tind},\gridpowvec]
{\cov}\inv[\gridpowvec]
\cov[\gridpowvec,\measpow\tnot{\tind}]\\
&=
\ushadvar
+\fadvar + \measnoisevar
-
(\covmat_{\shad(\loc\tnot{\tind}),\gridshadvec}
+
\covmat_{\fad(\loc\tnot{\tind}),\gridfadvec})\\&\quad\times
(\covmat_{\gridshadvec} + \fadvar \bm I_\gridnum)\inv
(\covmat_{\shad(\loc\tnot{\tind}),\gridshadvec}
+
\covmat_{\fad(\loc\tnot{\tind}),\gridfadvec})\transpose
\define\measlikelihoodvar\tnot{\tind},
\end{align*}
where the quantities
\begin{myitemize}%
\myitem\cmt{}$\updatelinearvec\tnot{\tind}\define
(\covmat_{\gridshadvec} + \fadvar \bm I_\gridnum)\inv
(\covmat_{\shad(\loc\tnot{\tind}),\gridshadvec}
+
\covmat_{\fad(\loc\tnot{\tind}),\gridfadvec})\transpose
$ and 
$ \updatelinearoffset\tnot{\tind}
\define
\basepow(\loc\tnot{\tind}) -
(\covmat_{\shad(\loc\tnot{\tind}),\gridshadvec}
+
\covmat_{\fad(\loc\tnot{\tind}),\gridfadvec})
(\covmat_{\gridshadvec} + \fadvar \bm I_\gridnum)\inv
\gridbasepowvec
$ have been defined along with
\myitem\cmt{}$\covmat_{\shad(\loc\tnot{\tind}),\gridshadvec}\define \cov[\shad(\loc\tnot{\tind}),\gridshadvec]$
and $\covmat_{\fad(\loc\tnot{\tind}),\gridfadvec}
\define\cov[\fad(\loc\tnot{\tind}),\gridfadvec]$. Clearly, the latter
contains a single non-zero entry if  $\loc\tnot{\tind}\in\grid$ and
vanishes otherwise.
\end{myitemize}%

\end{myitemize}%
\end{myitemize}%
\cmt{recombine}Finally, it follows from \cite[eq. (2.116)]{bishop2006} that
the requested posterior is
$p(\gridpowvec|\measpowvec\tnot{\tind})=\normal(\gridpowvec|
\meanvec_{\gridpowvec|\measpowvec\tnot{\tind}},\covmat_{\gridpowvec|\measpowvec\tnot{\tind}})$
with
\begin{align*}
\covmat_{\gridpowvec|\measpowvec\tnot{\tind}}&=( \covmat\inv_{\gridpowvec|\measpowvec\tnot{\tind-1}}+(1/\measlikelihoodvar\tnot{\tind})\updatelinearvec\tnot{\tind}
\updatelinearvec\tnot{\tind}\transpose)\inv\\
&=\covmat_{\gridpowvec|\measpowvec\tnot{\tind-1}}
-\frac{
 \covmat_{\gridpowvec|\measpowvec\tnot{\tind-1}}\updatelinearvec\tnot{\tind}
 \updatelinearvec\tnot{\tind}\transpose\covmat_{\gridpowvec|\measpowvec\tnot{\tind-1}}
}{
\measlikelihoodvar\tnot{\tind}+
\updatelinearvec\tnot{\tind}\transpose
\covmat_{\gridpowvec|\measpowvec\tnot{\tind-1}}
\updatelinearvec\tnot{\tind}
}
\\
\meanvec_{\gridpowvec|\measpowvec\tnot{\tind}}&=
\covmat_{\gridpowvec|\measpowvec\tnot{\tind}}\left[\frac{\measpow(\loc\tnot{\tind})-\updatelinearoffset\tnot{\tind}}{\measlikelihoodvar\tnot{\tind}}\updatelinearvec\tnot{\tind}
+\covmat\inv_{\gridpowvec|\measpowvec\tnot{\tind-1}}\meanvec_{\gridpowvec|\measpowvec\tnot{\tind-1}}\right].
\end{align*}
The sought algorithm applies these two update equations every time a
new measurement is acquired. The initializations are given by
$\covmat_{\gridpowvec|\measpowvec\tnot{-1}}
\define\covmat_{\gridshadvec} +\fadvar\bm I_\gridnum$ and
$\meanvec_{\gridpowvec|\measpowvec\tnot{-1}}\define \gridbasepowvec$.

\paragraph{Service Map Estimation}
\cmt{Overview}Since the service map $\serv(\loc)$ is a function of
$\pow(\loc)$, it is not surprising that the algorithm from the
previous section can be readily extended to obtain service maps. To
this end, apply Bayes rule and note that
$\serv(\gridloc\gridnot{\gridind})$ is deterministically solely
determined by $\pow(\gridloc\gridnot{\gridind})$ to write
\begin{myitemize}%
\myitem\cmt{marginal}
\begin{align*}
p(\serv(\gridloc\gridnot{\gridind})|\measpowvec\tnot{\tind})&=\int p(\serv(\gridloc\gridnot{\gridind}),\gridpowvec|\measpowvec\tnot{\tind})d \gridpowvec\\
&=\int p(\serv(\gridloc\gridnot{\gridind})|\gridpowvec ,\measpowvec\tnot{\tind})p(\gridpowvec |\measpowvec\tnot{\tind})d \gridpowvec\\
&=\int p(\serv(\gridloc\gridnot{\gridind})|\pow(\gridloc\gridnot{\gridind}) )p(\gridpowvec |\measpowvec\tnot{\tind})d \gridpowvec\\
&=\int p(\serv(\gridloc\gridnot{\gridind})|\pow(\gridloc\gridnot{\gridind}) )p(\pow(\gridloc\gridnot{\gridind} )|\measpowvec\tnot{\tind})d \pow(\gridloc\gridnot{\gridind}).
\end{align*}
Noting that $
p(\serv(\gridloc\gridnot{\gridind})|\pow(\gridloc\gridnot{\gridind})
)=1$ if $\pow(\gridloc\gridnot{\gridind})\geq \minpow$ and 0 otherwise,
the distribution of  $\serv(\gridloc\gridnot{\gridind})$
is fully characterized by 
\begin{align}
\servprob\gridnot{\gridind}\define \prob\left[\serv(\gridloc\gridnot{\gridind})=1|\measpowvec\tnot{\tind}\right]
&=\int_{\minpow}^{\infty}p(\pow(\gridloc\gridnot{\gridind} )|\measpowvec\tnot{\tind})d \pow(\gridloc\gridnot{\gridind}).
\end{align}
The latter expression can be evaluated through the cumulative
distribution function of a Gaussian random variable using the mean and
variance of $p(\pow(\gridloc\gridnot{\gridind}
)|\measpowvec\tnot{\tind})$ obtained earlier. 

\end{myitemize}%

\section{Adaptive Trajectory Design}
\label{sec:trajectory}

\nextv{
\cmt{Overview}The approach involves two steps: obtaining an uncertainty metric and a route through the locations with highest uncertainty.
\subsection{Trajectory Design Sub-Problem}
}
\begin{myitemize}%
\myitem\cmt{Outline}Since the UAV can navigate
to arbitrary locations in $\region$ to acquire measurements, the
problem becomes how to design a trajectory such that this acquisition
is performed as efficiently as possible.
\myitem\cmt{tradeoff}Since there is a trade-off between time and
estimation performance, a more formal problem statement would be, as
described later, to minimize the time required to obtain a map
estimate with a prescribed accuracy.
\myitem\cmt{accuracy}However, quantifying accuracy is itself a problem
since the true map is not available to the UAV. For this reason, it is
necessary to develop a suitable metric that the UAV can compute given
the measurements and prior information.
\end{myitemize}%

\subsection{Uncertainty Metric}
\label{sec:uncertainty}
\cmt{Uncertainty metric}The goal of this section is to design
$\uncert\gridnot{\gridind}(\measpowvec\tnot{\tind})\in [0,1]$, which
denotes the uncertainty in the target (power or service) map at
$\gridloc\gridnot{\gridind}$ after observing
$\measpowvec\tnot{\tind}$.
\nextv{and let
$\uncertvec(\measpowvec\tnot{\tind})\define[\uncert\gridnot{0}(\measpowvec\tnot{\tind}),\ldots,\uncert\gridnot{\gridnum-1}(\measpowvec\tnot{\tind})]\transpose$.
}%
\begin{myitemize}%
\myitem\cmt{def}%
\begin{myitemize}%
\myitem\cmt{power maps}If the goal is to estimate a power map, it seems reasonable
to use the posterior variance. To ensure that the resulting metric is
in $[0,1]$, one may normalize by the prior variance, since the latter
constitutes an upper bound for the posterior variance. This yields
\begin{align}
\label{eq:poweruncertainty}
\uncert\gridnot{\gridind}(\measpowvec\tnot{\tind})
= \frac{[\covmat_{\gridpowvec|\measpowvec\tnot{\tind}}]_{\gridind,\gridind}}{
\ushadvar + \fadvar}.
\end{align}
\myitem\cmt{service maps}In turn, for service map estimation, note that there is
little uncertainty when $\pow(\loc)$ is known to be very large or very
small: the most uncertain points are those where
$\servprob\gridnot{\gridind}$ is close to $1/2$. This is naturally
quantified by the posterior entropy of
$\serv(\gridloc\gridnot{\gridind})$:
\begin{align}
\label{eq:serviceuncertainty}
\uncert\gridnot{\gridind}(\measpowvec\tnot{\tind})
= -\servprob\gridnot{\gridind}\log_2(\servprob\gridnot{\gridind})
-(1-\servprob\gridnot{\gridind})\log_2(1-\servprob\gridnot{\gridind}).
\end{align}
\end{myitemize}%
\myitem\cmt{multiple channels}When there are multiple transmitters,
the values of the relevant metric (either \eqref{eq:poweruncertainty}
or \eqref{eq:serviceuncertainty}) for all transmitters can be
aggregated (e.g. by averaging or taking the maximum) to obtain a
single $\uncert\gridnot{\gridind}(\measpowvec\tnot{\tind})$ per
$\gridind$.

\myitem\cmt{Total}With these point-wise uncertainty metrics, one
can quantify the total uncertainty of the map after observing
$\measpowvec\tnot{\tind}$ by the spatial average
$\uncert(\measpowvec\tnot{\tind})\define (1/\gridnum)
\sum_{\gridind=0}^{\gridnum-1} \uncert\gridnot{\gridind}(\measpowvec\tnot{\tind})$.

\end{myitemize}%

\subsection{Route Planning}
\label{sec:routeplanning}
\begin{myitemize}%
\myitem\cmt{Policy}The UAV may use past measurements as well as prior
information about the map to decide where to measure next. Formally,
$\loc\tnot{\tind+1}=\policyfun(\measpowvec\tnot{\tind}, \locmat\tnot{\tind})$,
where function $\policyfun$ is the policy that needs to be designed.
\myitem\cmt{Cost}%
\begin{myitemize}
\myitem\cmt{Informal}Informally, one would like that
$\uncert(\measpowvec\tnot{\tind})$ decreases as fast as possible over
time. However, the specific criterion adopted to design $\policyfun$
may depend on the user's preferences.
\myitem\cmt{Formal}Let $\tfun(\locmat\tnot{\tind})$ denote  the time
that the UAV needs to follow the trajectory defined by the points in
$\locmat\tnot{\tind}$. A reasonable simplification is that the UAV
moves at constant speed $v$ and, therefore,
$\tfun(\locmat\tnot{\tind})
= \sum_{\auxtind=1}^\tind\|\loc\tnot{\auxtind}-\loc\tnot{\auxtind-1}\|/v$.
\begin{myitemize}%
\myitem\cmt{min cost}One may be, for example, interested in the $\policyfun$ that
minimizes $\expected[\uncert(\measpowvec\tnot{\tind})]$, where $\tind$ and
$\locmat\tnot{\tind}$ are such that $\tfun(\locmat\tnot{\tind})$ is
below a given upper bound.
\myitem\cmt{min time}Alternatively, one could minimize
$\expected[\tfun(\locmat\tnot{\tind})]$ subject to an upper bound on
$\uncert(\measpowvec\tnot{\tind})$.
\myitem\cmt{discounted uncertainty decrease}Yet another possible
criterion would be to maximize the discounted reward
$\expected[\sum_{\auxtind=1}^\tind \discount^{\tfun(\locmat\tnot{\auxtind})}(\uncert(\measpowvec\tnot{\auxtind-1})
-\uncert(\measpowvec\tnot{\auxtind}))]$ with $\discount\in(0,1)$
given. Clearly, this objective promotes trajectories with large uncertainty
improvements $\uncert(\measpowvec\tnot{\auxtind-1})
-\uncert(\measpowvec\tnot{\auxtind})$ at the beginning.
\end{myitemize}%
\end{myitemize}%

\myitem\cmt{NP-hardness}All these formulations
lead to non-convex optimization problems where the optimization
variable is the function $\policyfun$. Thus, it is necessary to
discretize the set of candidate measurement locations, for instance
by restricting $\loc\tnot{\tind}\in\grid$. Unfortunately, even in that
case,  this kind of problems can be shown to be NP-hard; see
e.g.~\cite{leny2009trajectory,blum2007orienteering} and references
therein. Thus, one needs to resort to approximations.

\begin{figure*}[t]
 \centering
 \includegraphics[width=1\textwidth]{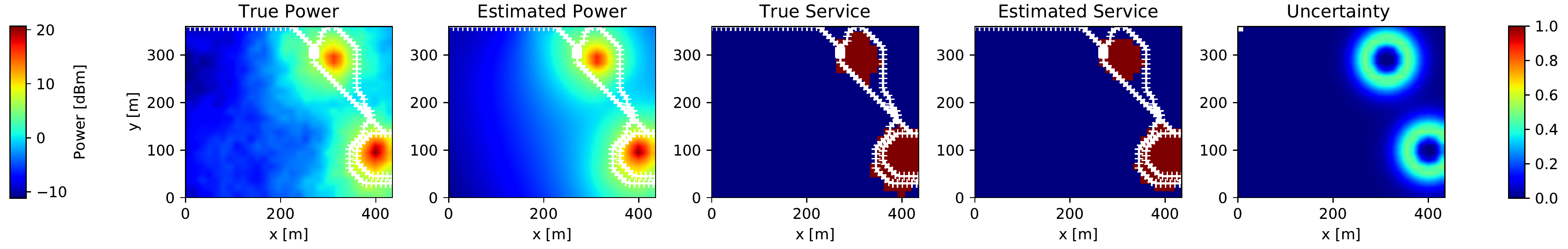}
 \caption{Trajectory (white line) followed by the autonomous UAV in a sample spectrum surveying operation. }
 \label{fig:realization}
\end{figure*}

\begin{figure}[t]
 \centering
 \includegraphics[width=.5\textwidth]{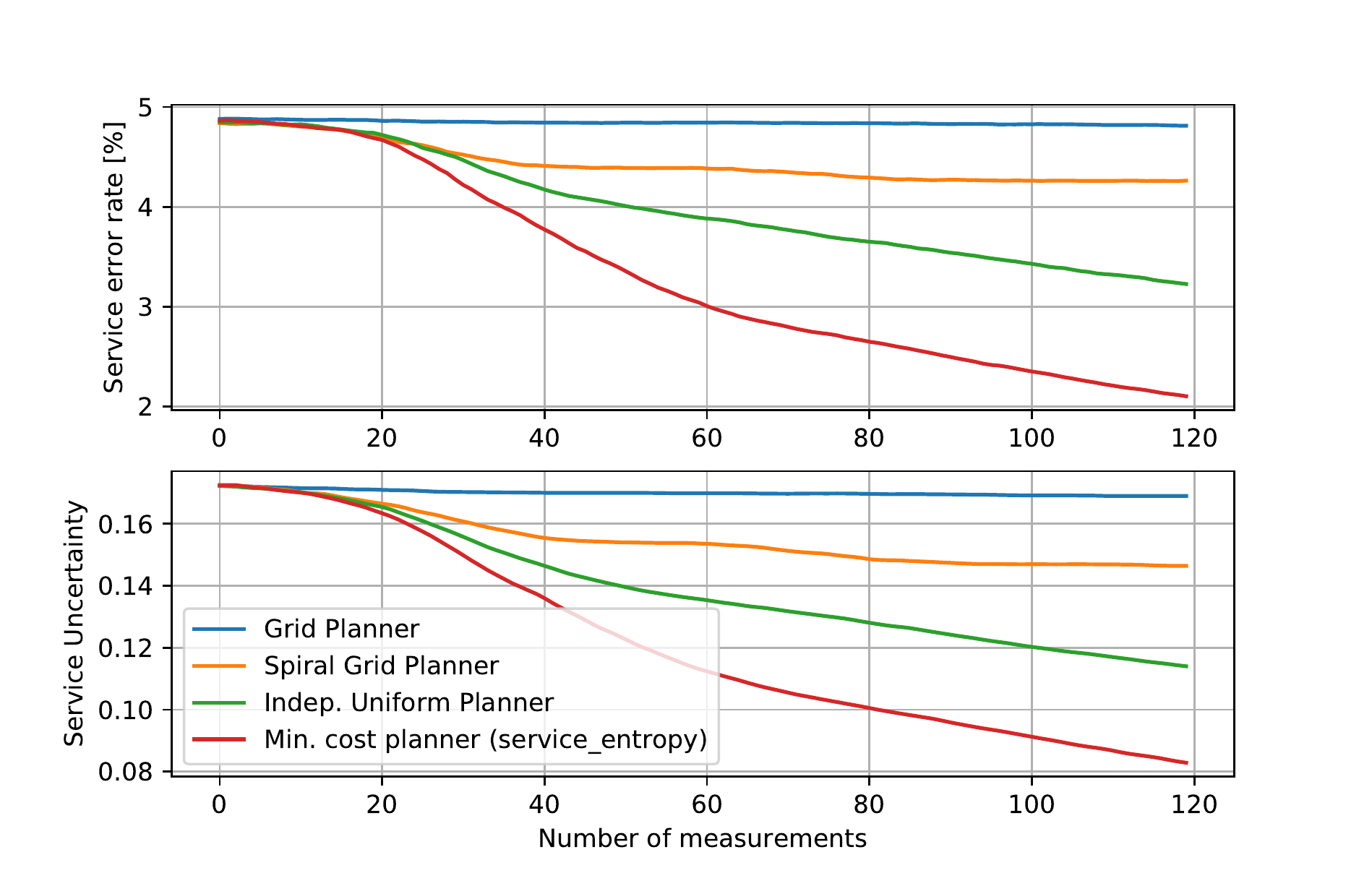}
 \caption{Comparison between the proposed minimum cost planner and
 three benchmarks. }
 \label{fig:uncertainty}
\end{figure}

\myitem\cmt{Approximations}%
\begin{myitemize}
\myitem\cmt{Power maps}In the case of power maps, note
that \eqref{eq:poweruncertainty} does not depend on the measurements,
but only on their location. Therefore, a (suboptimal) trajectory can
be found in an offline fashion, for example along the lines
of the algorithm in \cite{leny2009trajectory} and references therein.
\myitem\cmt{Service maps}In turn, for service maps, the metric \eqref{eq:serviceuncertainty}
does depend on the measurements and, therefore, the trajectory should
be computed on-the-fly, as measurements are collected.
\begin{myitemize}
\myitem\cmt{Receding horizon}However, updating the trajectory with
the reception of every new measurement may be too costly. Besides, the
presence of expectations in the aforementioned objectives renders such
a task intractable. A more sensible alternative is to update the
trajectory every $\tupdate$ measurements, assuming that
$\uncert\gridnot{\gridind}(\measpowvec\tnot{\tind})$ remains
approximately constant between consecutive updates at all grid points
except where a measurement is collected, in which case
$\uncert\gridnot{\gridind}(\measpowvec\tnot{\tind})$ becomes 0 at that
point. In other words,  measuring at location
$\loc\tnot{\tind}=\gridloc\gridnot{\gridind}$ yields $\uncert(\measpowvec\tnot{\tind})\approx
\uncert\gridnot{\gridind}(\measpowvec\tnot{\tind-1})-(1/\gridnum)
\uncert\gridnot{\gridind}(\measpowvec\tnot{\tind-1})
$.

\myitem\cmt{weighted-reward TSP}Such a receding horizon approach could
be cast as an instance of the so-called \emph{weighted-reward traveling salesman
problem} 
and a solution could be approximated by means of the algorithm
in \cite{blum2007orienteering}, which has a polynomial complexity.
\myitem\cmt{low-complexity}For real-time UAV operations, limited by
computational power, it may be preferable to pursue alternatives with
lower complexity, yet higher suboptimality. The alternative explored
here is to select, at each trajectory update, a destination in $\grid$
with highest local uncertainty. For rectangular $\grid$, if
$\uncert\gridnot{\gridind}(\measpow\tnot{\tind})$ is organized as a
matrix, this destination can be found as the maximum of such a matrix
spatially filtered by a low-pass kernel. The route to reach that
destination can be sought by minimizing the line integral of
$\uncert_{\loc}(\measpowvec\tnot{\tind})$ along the trajectory, where
$\uncert_{\loc}(\measpowvec\tnot{\tind})$ denotes the uncertainty at
$\loc$. This trajectory can be approximated through a
shortest-path algorithm (e.g. Bellman-Ford) with edge cost between
$\gridloc\gridnot{\gridind}$ and $\gridloc\gridnot{\gridind'}$ given
by the reciprocal of
$\int_{\gridloc\gridnot{\gridind}}^{\gridloc\gridnot{\gridind'}}\uncert_{\loc}(\measpowvec\tnot{\tind})d\loc\approx
\|{\gridloc\gridnot{\gridind'}}-{\gridloc\gridnot{\gridind}}\|(\uncert\gridnot{\gridind'}(\measpowvec\tnot{\tind})
-\uncert\gridnot{\gridind}(\measpowvec\tnot{\tind}))/2$. This clearly promotes paths through locations with
 high uncertainty. The trajectory can be recomputed periodically or
 after reaching each destination.  Although the resulting complexity
 is very low, the limitation is that wiggly trajectories, sometimes
 preferable~\cite{blum2007orienteering}, are penalized by this criterion.
\end{myitemize}%

\end{myitemize}%

\end{myitemize}%

\section{Numerical Experiments}
\label{sec:experiments}

\cmt{Overview}This section assesses the performance of the proposed
 algorithms by means of simulations. To ensure reproducibility, all
 the code will be made available at the authors' websites.

\cmt{simulation setup}
\begin{myitemize}%
\myitem\cmt{area}For simplicity, simulations are carried out assuming
 that the UAV stays at a constant height of $20$ m and, therefore,
 $\regiondim=2$. A $30 \times 25$ rectangular grid is constructed over
 the area of interest with a separation of $10$ m
 between each pair of adjacent grid points. 
\myitem\cmt{data generation}Two transmitters of height $10$ m  are
 deployed at locations drawn uniformly at
 random over $\region$.  The true map is generated by drawing $\gridpowvec$ from a
Gaussian distribution according to \eqref{eq:gridpowvec}, where
$\basepow(\loc)$ is obtained for a path loss exponent of 2, frequency
$2.4$~GHz, and isotropic transmit antennas. The transmit power is set
to $\txpow=10$ dBm for both sources.  Due to space limitations, we
focus on illustrating the effect of shadowing and, thus,
$\fadvar$ and $\measnoisevar$ are set to 0. The shadowing is generated
with $\dist_0=50$ m, $\ushadvar=9$, and $\ushadmean=0$. To generate
 measurements off the grid,  $\gridpowvec$  is interpolated using
cubic splines. To generate the service map, $\minpow$ is set to 5 dBm.

\myitem\cmt{tested algorithms}
\begin{myitemize}
\myitem\cmt{Proposed}The route planning algorithm
described at the end of Sec.~\ref{sec:routeplanning} is implemented
with a $3 \times 3$ kernel of all ones. A trajectory is updated only
every time the UAV reaches the destination. This update is performed
through the well-known Bellman-Ford algorithm for shortest path. The
candidate waypoints lie on $\grid$ and the UAV is allowed to move
in one out of 8 directions that differ 45 degrees. The uncertainty of the maps
corresponding to each transmitter is aggregated through a $\max$
operation; cf. Sec.~\ref{sec:uncertainty}.
\myitem\cmt{benchmarks}Since there is no algorithm for spectrum
surveying in the literature, the proposed method is compared against
three benchmarks. Each benchmark corresponds to a different approach
to plan the trajectory. The first follows parallel lines, thus having
waypoints on a rectangular grid (grid planner); the second follows a
rectangular spiral, and the third selects the next destination
uniformly at random, then moves there straight ahead.  To ensure a
fair comparison, all tested approaches collect a measurement every 5 m
on their trajectory. This means that, under the assumption of constant
speed, the time required to collect $\tind$ measurements is the same
for all approaches. Similarly, all approaches use the proposed
online estimator.
 
\end{myitemize}

\myitem\cmt{performance metrics}Performance is assessed in terms of
total uncertainty $\uncert(\measpowvec\tnot{\tind})$ and the service
error rate, which is the fraction of grid points
$\gridloc\gridnot{\gridind}$ where $\serv(\gridloc\gridnot{\gridind})$
differs from its estimate.
\end{myitemize}%
\cmt{description of the experiments}%
\begin{myitemize}%
\myitem\cmt{Maps}Fig.~\ref{fig:realization} depicts the true and
estimated power map, the true and estimated service map, as well as
the service uncertainty  (cf.~\eqref{eq:serviceuncertainty}) before starting to measure. As
observed, the
regions with highest uncertainty form rings around the sources (see
Sec.~\ref{sec:uncertainty}). The trajectory generated by the proposed
route planner is observed to target precisely these
locations.
\myitem\cmt{Curves}Fig.~\ref{fig:uncertainty} compares the reduction
of the error rate and uncertainty for the tested
algorithms with a Monte Carlo simulation. It is observed that the
proposed scheme results in a steeper
slope, confirming that it learns the map faster than the benchmarks.
\end{myitemize}

\section{Conclusions}

This paper proposes  collecting radio measurements with an autonomous
UAV to construct  radio power and service maps in two steps. First, an
online Bayesian learning algorithm obtains the posterior distribution
of the radio map at a set of grid locations given all past
measurements. This not only provides map estimates but also their
associated uncertainty via the posterior variance. Second, a route
planner algorithm uses the relevant uncertainty metric to plan a
trajectory along areas with  high uncertainty, which naturally leads
to acquire measurements at approximately the most informative locations. 

\printmybibliography

\begin{thebibliography}{10}

\bibitem{grimoud2010rem}
S.~Grimoud, S.~B. Jemaa, B.~Sayrac, and E.~Moulines,
\newblock ``A {REM} enabled soft frequency reuse scheme,''
\newblock in {\em Proc. IEEE Global Commun. Conf.}, Miami, FL, Dec. 2010, pp.
  819--823.

\bibitem{yilmaz2013radio}
H.~B. Yilmaz, T.~Tugcu, F.~Alag{\"o}z, and S.~Bayhan,
\newblock ``Radio environment map as enabler for practical cognitive radio
  networks,''
\newblock {\em IEEE Commun. Mag.}, vol. 51, no. 12, pp. 162--169, Dec. 2013.

\bibitem{chen2017map}
J.~Chen and D.~Gesbert,
\newblock ``Optimal positioning of flying relays for wireless networks: {A}
  {LOS} map approach,''
\newblock in {\em Proc. IEEE Int. Conf. Commun.} IEEE, 2017, pp. 1--6.

\bibitem{zhang2019constrained}
S.~Zhang, Y.~Zeng, and R.~Zhang,
\newblock ``Cellular-enabled {UAV} communication: A connectivity-constrained
  trajectory optimization perspective,''
\newblock {\em IEEE Trans. Commun.}, vol. 67, no. 3, pp. 2580--2604, Mar. 2018.

\bibitem{alayafeki2008cartography}
A.~Alaya-Feki, S.~B. Jemaa, B.~Sayrac, P.~Houze, and E.~Moulines,
\newblock ``Informed spectrum usage in cognitive radio networks: Interference
  cartography,''
\newblock in {\em Proc. IEEE Int. Symp. Personal, Indoor Mobile Radio Commun.},
  Cannes, France, Sep. 2008, pp. 1--5.

\bibitem{jayawickrama2013compressive}
B.~A. Jayawickrama, E.~Dutkiewicz, I.~Oppermann, G.~Fang, and J.~Ding,
\newblock ``Improved performance of spectrum cartography based on compressive
  sensing in cognitive radio networks,''
\newblock in {\em Proc. IEEE Int. Commun. Conf.}, Budapest, Hungary, Jun. 2013,
  pp. 5657--5661.

\bibitem{bazerque2010sparsity}
J.-A. Bazerque and G.~B. Giannakis,
\newblock ``Distributed spectrum sensing for cognitive radio networks by
  exploiting sparsity,''
\newblock {\em IEEE Trans. Signal Process.}, vol. 58, no. 3, pp. 1847--1862,
  Mar. 2010.

\bibitem{kim2013dictionary}
S.-J. Kim and G.~B. Giannakis,
\newblock ``Cognitive radio spectrum prediction using dictionary learning,''
\newblock in {\em Proc. IEEE Global Commun. Conf.}, Atlanta, GA, Dec. 2013, pp.
  3206 -- 3211.

\bibitem{lee2016lowrank}
D.~Lee, S.-J. Kim, and G.~B. Giannakis,
\newblock ``Channel gain cartography for cognitive radios leveraging low rank
  and sparsity,''
\newblock {\em IEEE Trans. Wireless Commun.}, vol. 16, no. 9, pp. 5953--5966,
  Jun. 2017.

\bibitem{tang2016spectrum}
M.~Tang, G.~Ding, Q.~Wu, Z.~Xue, and T.~A. Tsiftsis,
\newblock ``A joint tensor completion and prediction scheme for
  multi-dimensional spectrum map construction,''
\newblock {\em IEEE Access}, vol. 4, pp. 8044--8052, Nov. 2016.

\bibitem{huang2015cooperative}
D.-H. Huang, S.-H. Wu, W.-R. Wu, and P.-H. Wang,
\newblock ``Cooperative radio source positioning and power map reconstruction:
  A sparse {B}ayesian learning approach,''
\newblock {\em IEEE Trans. Veh. Technol.}, vol. 64, no. 6, pp. 2318--2332, Aug.
  2014.

\bibitem{teganya2019locationfree}
Y.~Teganya, D.~Romero, L.~M. Lopez-Ramos, and B.~Beferull-Lozano,
\newblock ``Location-free spectrum cartography,''
\newblock {\em IEEE Trans. Signal Process.}, vol. 67, no. 15, pp. 4013--4026,
  Aug. 2019.

\bibitem{romero2017spectrummaps}
D.~Romero, S-J. Kim, G.~B. Giannakis, and R.~L\'opez-Valcarce,
\newblock ``Learning power spectrum maps from quantized power measurements,''
\newblock {\em IEEE Trans. Signal Process.}, vol. 65, no. 10, pp. 2547--2560,
  May 2017.

\bibitem{romero2018blind}
D.~Romero, Donghoon Lee, and G.~B. Giannakis,
\newblock ``Blind radio tomography,''
\newblock {\em IEEE Trans. Signal Process.}, vol. 66, no. 8, pp. 2055--2069,
  2018.

\bibitem{bazerque2011splines}
J.-A. Bazerque, G.~Mateos, and G.~B. Giannakis,
\newblock ``Group-lasso on splines for spectrum cartography,''
\newblock {\em IEEE Trans. Signal Process.}, vol. 59, no. 10, pp. 4648--4663,
  Oct. 2011.

\bibitem{han2020power}
X.~Han, L.~Xue, F.~Shao, and Y.~Xu,
\newblock ``A power spectrum maps estimation algorithm based on generative
  adversarial networks for underlay cognitive radio networks,''
\newblock {\em Sensors}, vol. 20, no. 1, pp. 311, Jan. 2020.

\bibitem{teganya2020autoencoders}
Y.~Teganya and D.~Romero,
\newblock ``Data-driven spectrum cartography via deep completion
  autoencoders,''
\newblock in {\em IEEE Int. Conf. Commun., Jun. 2020, arXiv:1911.12810}.

\bibitem{chen2017segmented}
J.~Chen, U.~Yatnalli, and D.~Gesbert,
\newblock ``Learning radio maps for {UAV}-aided wireless networks: A segmented
  regression approach,''
\newblock in {\em Proc. IEEE Int. Conf. Commun.}, Paris, France, May 2017, pp.
  1--6.

\bibitem{zheng2020simultaneous}
Y.~Zeng, X.~Xu, S.~Jin, and R.~Zhang,
\newblock ``Simultaneous navigation and radio mapping for cellular-connected
  {UAV} with deep reinforcement learning,''
\newblock {\em available at arXiv::2003.07574}, 2020.

\bibitem{blum2007orienteering}
A.~Blum, S.~Chawla, D.~R. Karger, T.~Lane, A.~Meyerson, and M.~Minkoff,
\newblock ``Approximation algorithms for orienteering and discounted-reward
  {TSP},''
\newblock {\em SIAM J. Comput.}, vol. 37, no. 2, pp. 653--670, Jun. 2007.

\bibitem{gudmundson1991correlation}
M.~Gudmundson,
\newblock ``Correlation model for shadow fading in mobile radio systems,''
\newblock {\em Electron. Letters}, vol. 27, no. 23, pp. 2145--2146, 1991.

\bibitem{bishop2006}
C.~M. Bishop,
\newblock {\em Pattern Recognition and Machine Learning},
\newblock Information Science and Statistics. Springer, 2006.

\bibitem{kay1}
S.~M. Kay,
\newblock {\em Fundamentals of Statistical Signal Processing, {V}ol. {I}:
  Estimation Theory},
\newblock Prentice-Hall, 1993.

\bibitem{leny2009trajectory}
J.~Le~Ny and G.~J. Pappas,
\newblock ``On trajectory optimization for active sensing in gaussian process
  models,''
\newblock in {\em Proc. IEEE Conf. Decision Control/Chinese Control Conf.},
  Shanghai, China, Dec. 2009, pp. 6286--6292.

\end{thebibliography}
\end{document}